\documentstyle[11pt,paspconf,psfig]{article}
\begin{document}

\title{ Small Group Dynamics and Extended Gas}
\author{F. Combes}
\affil{DEMIRM, Observatoire de Paris, 61 Av. de l'Observatoire,
F-75 014, Paris, France}

\begin{abstract}
Interactions between galaxies have spectacular effects on
gas dynamics, and small galaxy groups are a privileged place
to investigate them. In particular, they could test the 
existence of cold H$_2$ gas as dark matter in the outer
parts of galaxies.
 HI observations have revealed that galaxies in small groups 
are deficient in atomic gas, like in richer galaxy
clusters such as Virgo, although in a lesser extent.
Galaxy interactions could be the cause of this deficiency,
stripping the gas out of galaxies and enriching the
inter-cluster medium (ICM) in hot gas, which in turn strips
gas through ram pressure.  Alternatively, the gas present
at the formation of the group could have been heated to 
its virial temperature, and be observed now
as X-rays. The dynamical processes related
to this extended gas in small galaxy groups are reviewed. 
\end{abstract}


\vspace*{-0.3cm}
\section{Introduction}

Interaction of galaxies is a way to reveal extended gas
around galaxies, that could stay cold and dark in 
quiescent phases. The hypothesis that the
baryonic dark matter around galaxies could be 
cold  H$_2$ clouds has been developped in recent
years, in different models. The cold clouds could be
rotationally supported, 
orbiting in the outer parts of the galaxies
in a flaring plane, associated to the HI gas
(Pfenniger et al 1994, Pfenniger \& Combes 1994).
They are distributed in a fractal structure, with
the smallest fragments (the clumpuscules) having a mass 
M $\sim 10^{-3}$ M$_\odot$, a molecular
density n$_{\rm H2}\sim  10^{10}$ cm$^{-3}$, and a radius
R $\sim$ 10 AU. 
Alternatively, the cold H$_2$ gas could be associated 
to brown dwarfs in clusters, distributed
in the galactic halo, or spheroidal component
(de Paolis et al 1995; Gerhard \& Silk 1996).
Their temperature would then be higher (T= 20K).

These clumps are seen in ESE (extreme scattering events) in
front of quasars and in HI- VLBI in absorption against extended
radio-sources 
(Fiedler et al 1987, Walker \& Wardle 1998, Faison et al 1998).
The statistics of observations are compatible with 
a large number of these clumps in the Galaxy, so that
the total mass could be comparable to that of the visible
Milky Way. 

The existence of this cold H$_2$ gas has also been 
invoked to explain the extra-$\gamma$ rays
detected by EGRET, on board GRO 
(de Paolis et al. 1999, Sciama 1999)
and some observational consequences have been
studied (Combes \& Pfenniger 1997, Shchekinov 1999). 
Cooling flows in clusters could also form 
these clumpuscules (Ferland et al 1994).

Let us recall that the visible matter in the Universe 
corresponds to only $\Omega_{vis} \sim$ 0.003,
while the Big Bang nucleosynthesis implies that 
the baryonic density is $\Omega_b \sim$ 0.01 $h^{-2}$,
or that $ 0.01 < \Omega_b < 0.04$. Most of the baryons 
are therefore dark, and the MACHOs cannot be a 
significant part of them, according to the microlensing
experiments.

\section{Gas Dynamics in Small Groups}

\subsection{Atomic Gas}

The most striking feature in interacting galaxy groups
is the presence of large tidal tails of matter dragged out of
the galaxies, huge HI extensions with respect to the optical systems.
Yun et al. (1993) have found large quantities
of HI all around the M81/M82/NGC 3077 system, and Hibbard (1995) shows
in his thesis an evolving sequence of interacting/merging galaxies,
where the HI extensions are conspicuous. More precisely, the percentage 
of the total HI found in the tails/extensions is increasing with
the merging stage, from 20\% in the M81 system, to 80\% in the
merger remnant NGC 7252. This does not mean that most of the gas
will be expelled from interacting systems. In fact, with all probability, the gas 
dragged out remains bound to the system, and will rain back onto the merger
remnant, after some billion years. Hibbard et al. (1994) show that the
gas at the bottom of the tails in NGC 7252 is infalling. 

In normal galaxies, large HI extents are very rare
(Briggs et al 1980, Broeils 1992, Hoffman et al 1996).
In average the gas extent is  R$_{\rm HI} \sim$ 2 R$_{opt}$.
Only interacting systems show extended gas.

\subsection{Molecular Gas}

What is seen in the molecular phase is just the contrary: apparently 
large H$_2$ concentrations pile up at the galaxy nuclei in interacting
systems. Up to 50\% of the dynamical mass could be under
the form of molecular hydrogen in merging systems (Scoville et al. 1991),
even when the the  CO/H$_2$ conversion ratio is corrected to a minimum
(Bryant \& Scoville 1999, Solomon et al 1997).

In summary, the observations suggest that the HI gas is dragged outwards,
while the H$_2$ gas is driven inwards, to be consumed in star formation.
In fact these two tracers (HI and CO) shed light on two aspects of
the same gas component. In normal spiral galaxies,
there exists a sharp cut-off in the HI line distribution (21cm)
(van Gorkom et al 1991, Corbelli \& Salpeter 1993);
it is interpreted as a ionizing front, due to photoionisation from 
the extragalactic background. HII is the third gas phase 
to take into account.

While star formation is enhanced in interacting galaxies, this is not
true in groups and clusters.
This is mainly due to the stripping of the outer gas
acting as reservoirs of fuel for star formation.
The HI deficiency in compact groups is in average 3
(Williams \& Rood 1987, Williams et al 1991, Huchtmeier 1997,
Oosterloo \& Iovino 1997),
and there is no significant CO deficiency or enhancement
(Boselli et al 1996, Leon et al 1997, Verdes-Montenegro et al. 1998).
The missing gas is heated by repeated interactions, 
and joins the coronal hot phase.
ROSAT survey of 22 HCG (Ponman et al 1996) have revealed that
75\% possess a diffuse hot gas. There appears to be a
correlation between X-ray and spiral galaxy fraction
(Pildis et al. 1995). 

\begin{figure}[t]
\centerline{
\psfig{figure=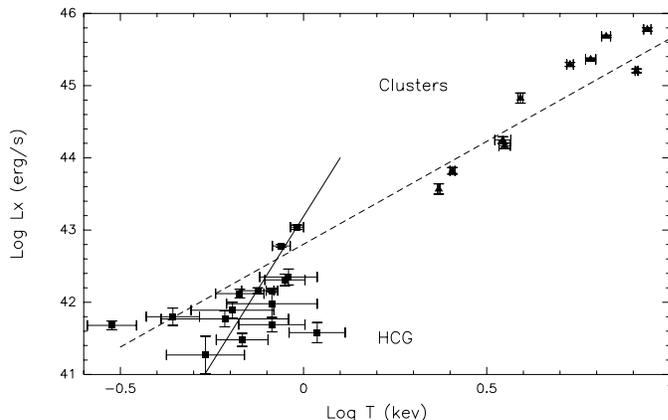,bbllx=2cm,bblly=0cm,bburx=18cm,bbury=26cm,width=9cm,angle=-90}
}
\caption{L$_x$ vs T for HCG (filled squares),
and clusters (filled triangles), from Ponman et al. (1996). The dashed line is the best
regression fit for clusters, and full line for HCGs. }
\label{lx}
\end{figure}

\subsection{Hot Gas }

Hot gas and its X-ray emission are the keys
to follow the fate of gas in interactions.
Read \& Ponman (1998) have followed
a merger sequence and found that the X-ray 
luminosity is enhanced by the interaction,
although not as much as the FIR luminosity.
Spatially, the X-ray emission  is not correlated with tidal tails;
the hot gas is ejected in winds, powered by the central
starbursts. The classical M82-type bipolar outflow
is often replaced by  unipolar outflows
in violent mergers.

In merger remnants (like in NGC 7252) L$_X$ falls down again.
This is surprising since such remnants are bound to become
classical ellipticals, which 
have more hot gas than spirals (e.g. Fabbiano, 1989).
How is done the transformation?
At the base of the tidal tails, in NGC 7252, the HI is falling back,
as revealed by its kinematics (Hibbard et al. 1994), but then
stops suddenly. It is not replaced by H$_2$ gas, that is 
concentrated much further in. It is likely that the
returning gas is heated by shocks to form a hot gas halo.
This will take a time-scale of the order of Gyrs to
form a genuine elliptical.

In small groups, the hot gas is even more extended, up to 400 kpc
(Davis et al. 1995). The amount of hot gas is such that
20 to 34\% of the mass is baryonic.
This cannot be done by galactic winds
(Ponman et al. 1996, Mulchaey et al. 1996).
About 75\% of compact groups have detectable 
X-ray emission (Ponman et al 1996). In small groups,
the hot gas metallicity is low (lower than in clusters).
In the L$_X$-T diagram, the compact groups occupy 
a special region, after a turn-over: they have less luminosity
for a given temperature than would be expected from
clusters (cf fig. \ref{lx}). This could be due to the
influence of galactic winds, creating a hole in the center,
while winds have no influence in clusters (Ponman et al 1996).
Alternatively, the hot gas could be progressively heated, 
through interactions, and have not yet reached equilibrium
in small groups. This would be the case if the hot gas comes
from the cold molecular gas in outer parts of galaxies
(Pfenniger \& Combes 1994).

\begin{figure}[t]
\centerline{
\psfig{figure=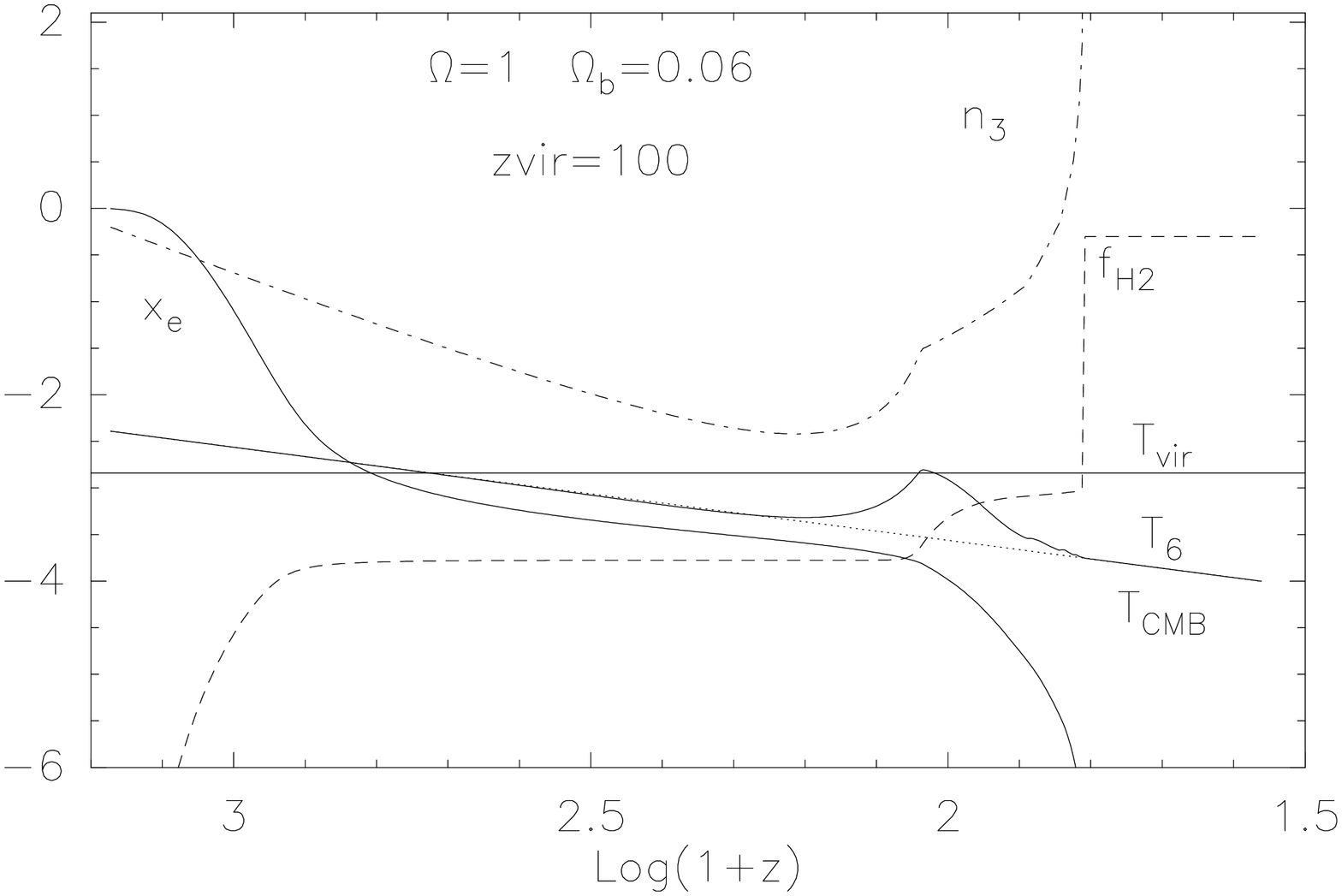,bbllx=2cm,bblly=2cm,bburx=18cm,bbury=26cm,width=11cm,angle=-90}
}
\caption[h]{ Evolution of temperature ($T_6$ in 10$^6$\,K), molecular
fraction ($f_{\rm H_2}$), ionization fraction ($x_{\rm e}$) and
density ($n_3$ in 10$^3$\,cm$^{-3}$) for a gas cloud after
recombination, collapsing at $z_{\rm vir} = 100$, with
$h=0.5$, $\Omega= 1$  and $\Omega_{\rm b}=0.06$ 
(from Combes \& Pfenniger 1998). }
\label{clump}
\end{figure}

\section{Formation of Gas Clumps}

In this hypothesis, what is the fate of the gas, and when do the
cold gas clumps form? After recombination ($z \sim 1500$), the largest masses
to get non-linear are of $M \sim 10^{6-8}$ M$_\odot$,
 and the Jeans mass is $M \sim 10^5$\,M$_\odot
({{\Omega_{\rm b}}\over{0.06}})^{-1/2} ({{h}\over{0.5}})^{-1}$.
All masses between these two will collapse and decouple from expansion,
but the structures at precisely the Jeans mass collapse first.  The
masses correspond to typical $z=0$ giant molecular clouds.
If cooling is efficient enough ($\tau_{\rm cool} \sim \tau_{\rm ff}$),
the collapse is quasi-isothermal, and fragmentation occurs, since the
Jeans mass becomes smaller and smaller
as the density increases (e.g., Hoyle 1953).  Fragmentation is
limited by opacity, and the smallest fragments (or clumpuscules),
which are at the transition of being 
pressure supported, are today of the order of $10^{-3}$ M$_\odot$, and
their mass grows slowly, as $T^{1/4}$ or $(1+z)^{1/4}$, with redshift.

Cooling might be a problem, since primordial gas is non-metallic,
with no dust grains;
above $10^4$\,K the main coolant is atomic hydrogen (by collisional
excitation of Ly$\alpha$), then 
vibration-rotation lines of molecular hydrogen take over until
$T=200$\,K, because a significant quantity of H$_2$ molecules 
forms through H$^-$ and H$_2^+$.  Below 200\,K, HD is then more
efficient. Many groups have computed 
the physico-chemistry of the primordial gas, 
to determine the size of the first forming bound structures (Yoneyama
1972; Hutchins 1976; Carlberg 1981; Palla et al.\ 1983; Lepp \& Shull
1984). All of them have found that the cooling is indeed efficient, as
soon as the redshift is below $z\sim 200$.

We have studied in particular the recursive formation of fragments, as expected
in a fractal structure of dimension D=1.7 (which fixes the density as
a function of sizes and masses). Cooling in such a fractal is very
efficient, as shown by the temperature curve, which quickly
returns back to T$_{CMB}$ of the background after virialisation 
(cf fig. \ref{clump}). Very soon, the clumps are almost entirely molecular
(Combes \& Pfenniger 1998).

\begin{figure}[t]
\centerline{
\psfig{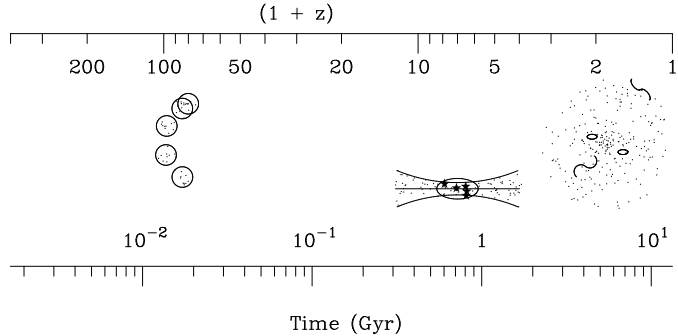}
}
\caption{ Schematic view of the evolution of the baryonic dark matter,
under the form of cold H$_2$ gas: formation of clumpuscules at a redshift
around 100, which are progressively involved in the formation of larger
structures and proto-galaxies. A typical early galaxy is shown at $ z = 6$,
with the cold gas settling into a flaring disk, and the star formation 
beginning in the center, where the surface density is above threshold.
Later on, when groups and clusters virialise, the gas is stripped and 
heated to contribute to the hot X-ray gas.}
\label{scen}
\end{figure}

\section{Discussion}

After this efficient formation, most of the gas is not
in favorable conditions for star formation. Indeed,
larger potential wells have not formed yet, and the
density threshold is not reached. It is therefore likely that the
gas will slowly accumulate in bigger potentials to form
galaxies, as sketched in fig. \ref{scen}. In the inner parts of
these early galaxies, star formation can then begin.
Then galaxy interactions will stirr and heat the gas through 
shocks and gravitational perturbations.

Besides, interactions accelerate the angular momentum 
transfer: part of the HI gas is dragged outwards in tails.
Most of the gas is driven inwards, giving rise to huge
nuclear starbursts (and may be AGN).
Galaxy evolution is highly accelerated.
The cold gas that was settled around each galaxy is heated 
and virialised in the new common potential and might be
visible through X-rays.


\end{document}